\documentclass[journal,oneside,final,letterpaper,twocolumn]{IEEEtran}%
\usepackage{amsmath}
\usepackage{amsfonts}
\usepackage{cite}
\usepackage{amssymb}
\usepackage{graphicx}%
\providecommand{\U}[1]{\protect\rule{.1in}{.1in}}
\providecommand{\U}[1]{\protect\rule{.1in}{.1in}}

\newtheorem{proposition}{Proposition}

\begin{document}

\title{Power Loading in Parallel Diversity Channels Based on Statistical Channel Information}
\author{Justin P. Coon,~\IEEEmembership{Senior~Member,~IEEE,} and Rafael Cepeda,
\IEEEmembership{Senior~Member,~IEEE}\thanks{J. P. Coon is with Toshiba
Research Europe Ltd., 32 Queen Square, Bristol, BS1 4ND, UK; tel: +44 (0)117
906 0700, fax: +44 (0)117 906 0701, email: justin@toshiba-trel.com.
\par
R. Cepeda was with Toshiba Research Europe Ltd. and is now with British Sky
Broadcasting Ltd., Grant Way, Isleworth, Middlesex, TW7 5QD, UK; tel: +44
(0)20 7032 7752, fax: +44 (0)20 7900 7161, email: rafael.cepeda@bskyb.com.}}
\maketitle

\begin{abstract}
In this paper, we show that there exists an arbitrary number of power
allocation schemes that achieve capacity in systems operating in parallel 
channels comprised of single-input multiple-output (SIMO) Nakagami-$m$ fading subchannels when the
number of degrees of freedom $L$ (e.g., the number of receive antennas)
tends to infinity. Statistical waterfilling -- i.e., waterfilling using
channel statistics rather than instantaneous channel knowledge -- is one such
scheme. We further prove that the convergence of statistical waterfilling to
the optimal power loading scheme is at least $\mathcal{O}\left(  1/\left(  L\log
L\right)  \right)  $, whereas convergence of other schemes is at worst
$\mathcal{O}\left(  1/\log L\right)  $.  To validate and demonstrate the practical use of our findings, 
we evaluate the mutual information of example SIMO parallel channels using simulations as well as 
new measured ultrawideband channel data.

\end{abstract}


\section{Introduction\label{sec:intro}}

Parallel channels are frequently encountered in modern communication systems.
Examples of such systems include those using orthogonal frequency division
multiplexed (OFDM) transmissions, time-division multiplexed (TDM)
transmissions, frequency-hop spread-spectrum (FH-SS), and multiple-input
multiple-output (MIMO) systems that employ eigenmode transmission techniques.

In recent years, the design of subchannel power allocation strategies has
become a popular research topic for systems operating in parallel channels.
Power loading schemes based on having perfect channel state information at the
transmitter (CSIT) have been developed to maximize spectral efficiency
\cite{Lei2004a} and system throughput \cite{Bansal2008}, as well as to
minimize bit-error rate~\cite{Goldfeld2002,Ermolova2007} and transmit power
\cite{Liu2005}. Solutions that rely on partial CSIT have also been
devised (see, e.g.,~\cite{Vu2007}).

Here, we investigate power loading in parallel channels using only
\emph{statistical} CSIT (SCSIT). Several researchers have reported notable
results on the use of SCSIT in the literature. For example, power loading
based on SCSIT to minimize bit-error rate was studied in \cite{Digham2004},
while in \cite{Song2002}, the authors developed power and bit allocation
algorithms to maximize the spectral efficiency and minimize the power
consumption. More recently, in~\cite{Gao2009}, the authors investigated
transmission in jointly correlated MIMO channels, whereby the transmit vector
is conveyed over the eigenmodes of the transmit correlation matrix, and power
allocation is performed to maximize an upper bound on the ergodic capacity. It
was found that, for parallel channels in particular, \emph{statistical
waterfilling} -- i.e., waterfilling using the \emph{mean} channel gains
instead of the \emph{instantaneous} gains -- optimized this bound \cite[cf.
(67)]{Gao2009}.

In this paper, we study power loading in parallel channels comprised of
single-input multiple-output (SIMO) subchannels where the number of degrees of
freedom of each subchannel (e.g., the number of receive antennas) is
denoted by $L$. We maintain generality in our analysis by assuming the
single-input single-output (SISO) channels that constitute each SIMO
subchannel adhere to a Nakagami-$m$ fading model. In contrast to
\cite{Gao2009}, we investigate conditions under which we can determine the
power loading schemes that maximize the \emph{exact} ergodic capacity of such
a parallel channel rather than an upper bound. To this end, we show that as
$L\rightarrow\infty$, there are arbitrarily many such schemes, where
statistical waterfilling is one such approach. Furthermore, we demonstrate
that convergence to the optimal power allocation scheme is $\mathcal{O}\left(
1/\left(  L\log L\right)  \right)  $ for statistical waterfilling, whereas the
convergence of other asymptotically optimal schemes is at worst $\mathcal{O}%
\left(  1/\log L\right)  $.  Finally, we validate and demonstrate the practical 
use of our findings through a study of SIMO parallel channels using simulations as well as 
new measured ultrawideband (UWB) channel data.

The rest of the paper is organized as follows. Bounds on the capacity of a
parallel channel are given in Section \ref{sec:bounds}. These bounds are used
to develop the main results on the asymptotic optimality and convergence
properties of statistical waterfilling and other power loading strategies in
SIMO systems in Section \ref{sec:main}. Measured and simulated channel data are used
to measure performance in Section \ref{sec:eval}, and conclusions are then drawn in Section
\ref{sec:conclusions}.

\section{Bounds on Capacity\label{sec:bounds}}

Consider a parallel channel comprised of $N$ SIMO subchannels, each with $L$
degrees of freedom\footnote{The restriction that all subchannels have $L$
degrees of freedom is made for simplicity of presentation; in fact, the main
result of this paper holds even if $L$ varies across the subchannels.}. The
mutual information\footnote{Mutual information and capacity are defined in
units of nats.} of the $n$th subchannel, assuming full channel state
information is available at the receiver, is given by \cite{Tse2005}%
\[
I_{\gamma_{n}}\left(  P_{n}\right)  =\log\left(  1+\frac{P_{n}}{N_{0}}%
\gamma_{n}\right)
\]
where $\gamma_{n}=\left\Vert \mathbf{h}_{n}\right\Vert ^{2}$ with
$\mathbf{h}_{n}\in\mathbb{C}^{L}$ being a vector of complex channel
coefficients, $P_{n}$ is the power transmitted on this subchannel, and $N_{0}$
is the variance of the zero-mean, additive white Gaussian noise on that
subchannel\footnote{We assume without loss of generality that the power
spectral densities of the noise processes on all subchannels are identical.}.
It follows that the ergodic capacity of the parallel channel is given by%
\begin{equation}
C=\sup_{\substack{P_{n}\geq0,\\\sum P_{n}=P}}\left\{  \sum_{n=1}^{N}E\left[
I_{\gamma_{n}}\left(  P_{n}\right)  \right]  \right\}  . \label{eq:CE}%
\end{equation}
Since only SCSIT is available, the powers $\left\{  P_{n}\right\}  $ cannot be
functions of $\left\{  \gamma_{n}\right\}  $, but instead rely upon the
statistics of these channel gains. We define the set $\left\{  P_{n}^{\star
}\right\}  $ as the set of optimal powers, i.e., the set that yields
the supremum given above.

To develop our main results, we will require an upper and a lower bound on
$C$. An upper bound is easily calculated using Jensen's inequality. Let
$P_{n}^{s}$ denote the power loading strategy based on statistical
waterfilling, which is given by%
\begin{equation}
P_{n}^{s}=\left(  \nu-\frac{N_{0}}{\mu_{n}}\right)  ^{+} \label{eq:swfill}%
\end{equation}
where $\left(  x\right)  ^{+}=\max\left\{  0,x\right\}  $, $\mu_{n}=E\left[
\gamma_{n}\right]  $ is the mean of the $n$th subchannel gain, and $\nu$ is
chosen to satisfy $\sum P_{n}^{s}=P$. Now, we can write%
\begin{equation}
C\leq\sum_{n=1}^{N}I_{\mu_{n}}\left(  P_{n}^{\star}\right)  \leq\sum_{n=1}%
^{N}I_{\mu_{n}}\left(  P_{n}^{s}\right)  \label{eq:ub1}%
\end{equation}
where the first inequality follows from the concavity of $I$ and the second
results from the standard waterfilling solution.

Clearly, $\left\{  P_{n}^{s}\right\}  $ is not the only set that satisfies
this bound; it is simply the set that \emph{maximizes} the bound. It will be
convenient to define another set of powers $\left\{  P_{n}^{\prime}\right\}  $
that satisfies%
\begin{equation}
C\leq\sum_{n=1}^{N}I_{\mu_{n}}\left(  P_{n}^{\prime}\right)  \leq\sum
_{n=1}^{N}I_{\mu_{n}}\left(  P_{n}^{s}\right)  \label{eq:ub2}%
\end{equation}
such that $\sum P_{n}^{\prime}=P$. Note, we do not explicity define $\left\{
P_{n}^{\prime}\right\}  $.

For the lower bound, we begin with the trivial inequality%
\begin{equation}\label{eq:lb1}
C\geq\sum_{n=1}^{N}E\left[  I_{\gamma_{n}}\left(  P_{n}^{\prime}\right)
\right]  .
\end{equation}
Applying Markov's inequality yields the more useful bound%
\begin{equation}
C\geq\sum_{n=1}^{N}a_{n}\mathcal{F}_{n}\left(  \frac{N_{0}}{P_{n}^{\prime}%
}\left(  e^{a_{n}}-1\right)  \right)  \label{eq:lb2}%
\end{equation}
where $\mathcal{F}_{n}\left(  x\right)  =\Pr\left(  \gamma_{n}\geq x\right)  $
is the complementary cumulative distribution function (CCDF) of the $n$th
subchannel gain and $a_{n}>0$. Note that a sufficient (but not necessary)
condition for this bound to hold is $P_{n}^{\prime}=P_{n}^{s}$ for all $n$.

\section{Power Loading in Diversity Channels\label{sec:main}}

We now present our main results. We consider the case where each subchannel
experiences Nakagami-$m$ fading. The constituent single-input single-output
(SISO) channels corresponding to the $n$th SIMO subchannel are statistically
independent, and the $\ell$th SISO channel gain, denoted by $\left\vert
h_{n,\ell}\right\vert ^{2}$, is gamma distributed with scale parameter
$\theta_{n}$ and shape parameter $m_{n}$. Note that we assume the scale and
shape parameters are identical for all SISO channels that constitute the $n$th
subchannel. Thus, the channel gain $\gamma_{n}=\left\Vert \mathbf{h}%
_{n}\right\Vert ^{2}$ is a gamma distributed variate with scale parameter
$\theta_{n}$ and shape parameter $m_{n}L$. The density function of $\gamma
_{n}$ is given by%
\begin{equation}
f_{n}\left(  \gamma\right)  =\frac{\gamma^{m_{n}L-1}e^{-\frac{\gamma}%
{\theta_{n}}}}{\theta_{n}^{m_{n}L}\Gamma\left(  m_{n}L\right)  },\quad
\gamma\geq0. \label{eq:mrc_density}%
\end{equation}
The mean of the $n$th subchannel gain in this case is given by%
\begin{equation}
\mu_{n}=\theta_{n}m_{n}L
\end{equation}
Now we can study the capacity as the number of degrees of freedom $L$ grows
large, which leads to the following proposition.

\begin{proposition}
\label{th:mrc_optimal_power}Consider a parallel channel with subchannel gains
distributed according to (\ref{eq:mrc_density}). In the limit of large $L$,
there are arbitrarily many asymptotically optimal power loading strategies,
one of which is the statistical waterfilling solution defined by
(\ref{eq:swfill}).
\end{proposition}

\begin{IEEEproof}
We prove the proposition by showing that the upper and lower bounds given by%
\begin{equation}
\sum_{n=1}^{N}E\left[  I_{\gamma_{n}}\left(  P_{n}^{\prime}\right)  \right]
\leq C\leq\sum_{n=1}^{N}I_{\mu_{n}}\left(  P_{n}^{\prime}\right)
\label{eq:ulb}%
\end{equation}
are asymptotically equivalent (as $L\rightarrow\infty$) for any set $\left\{
P_{n}^{\prime}\right\}  $ that satisfies~\eqref{eq:ulb} for $L > L_0$. Note that we
do not explicitly
define $\left\{  P_{n}^{\prime}\right\}  $, although it is clear that one possible definition is $P_n^{\prime} = P_n^s$ for all $n$ since this definition satisfies~\eqref{eq:ulb} for all $L$.
We first consider the upper and lower bounds on the capacity of the $n$th
subchannel. We assume that a nonzero power is allocated for transmission on
this subchannel; this is a valid assumption since if the converse were true,
the upper and lower bounds on capacity would both be zero. The CCDF of the
$n$th subchannel gain is%
\begin{equation}
\mathcal{F}_{n}\left(  x\right)  =Q\left(  m_{n}L,\frac{x}{\theta_{n}}\right)
\end{equation}
where $Q\left(  a,x\right)  =\Gamma\left(  a,x\right)  /\Gamma\left(
a\right)  $ is the normalized upper incomplete gamma function. Now, letting
$\beta_{n}=P_{n}^{\prime}\theta_{n}m_{n}/N_{0}$ where $P_{n}^{\prime}>0$, we can
write the ratio of the lower and upper bounds given by (\ref{eq:lb2}) and
(\ref{eq:ub2}), respectively, as a function of $L$:%
\begin{equation}
r_{L}=\frac{a_{n}Q\left(  m_{n}L,\beta_{n}^{-1}m_{n}\left(  e^{a_{n}%
}-1\right)  \right)  }{\log\left(  1+\beta_{n}L\right)  }.
\end{equation}
For every $L\in\mathbb{N}$ and $n\in\{1,2,\ldots,N\}$, we can calculate
$\beta_{n}$. Thus, we can choose to define%
\begin{equation}
a_{n}=\log\left(  1+\alpha \beta_{n} L\right)  >0
\end{equation}
where $0<\alpha<1$. It follows that%
\begin{equation}\label{eq:rL}
r_{L}=\frac{\log\left(  1+\alpha \beta_{n} L\right)  }{\log\left(  1+\beta
_{n}L\right)  }Q\left(  m_{n}L,\alpha m_{n}L\right)  .
\end{equation}
Using \cite[5.11.3]{NIST2010} and \cite[8.11.6]{NIST2010}, and noting that
$0<\alpha e^{1-\alpha}<1$ for $0<\alpha<1$, we can expand $r_{L}$ at $\infty$
to obtain%
\begin{align}
r_{L} &  =\underbrace{\left(  1+\frac{\log\alpha}{\log L}+\mathcal{O}\left(
\frac{1}{\left(  \log L\right)  ^{2}}\right)  \right)  }_{\text{logarithmic
term}}\nonumber\\
&  \quad\times\underbrace{\left(  1-\frac{\left(  \alpha e^{1-\alpha}\right)
^{m_{n}L}}{\left(  1-\alpha\right)  \sqrt{2\pi m_{n}L}}\left(  1+\mathcal{O}%
\left(  L^{-1}\right)  \right)  \right)  }_{\text{gamma function term}}.
\end{align}
Thus, $r_{L}\rightarrow1$ as $L\rightarrow\infty$. Since this relation holds
for any $n$, the upper and lower bounds given in
(\ref{eq:ulb}) are asymptotically equivalent, and $\left\{  P_{n}^{\prime} \right\}  $ is an optimal set of powers. Finally, it follows from
continuity that any set $\left\{  P_{n}^{\prime}+\epsilon_{n}\right\}  $ with
$\epsilon_{n}\in\mathbb{R}$, which satisfies~\eqref{eq:ulb}, is also optimal.
\end{IEEEproof}

This result can be understood intuitively by noting that the mean channel
gains increase monotonically with the number of degrees of freedom $L$. Thus,
the effective SNR on a given subchannel grows without bound as $L\rightarrow
\infty$, suggesting that an equal power allocation strategy is optimal. Many
power loading strategies satisfy this condition, including traditional
waterfilling and statistical waterfilling. In fact, this result is also
related to the notion of \textquotedblleft channel hardening\textquotedblright%
, which has been studied in the information theoretic literature. In particular, our result is corroborated by
related findings detailed recently in \cite[cf. \S IV]{Bai2009}.

Proposition \ref{th:mrc_optimal_power} is stronger than it may first appear
since it implies that one may
theoretically determine the number of diversity branches such that, when an
asymptotically optimal power loading strategy is employed, the ergodic
capacity can be approached to an arbitrarily close degree. More exactly, for
arbitrarily small $\epsilon>0$, there exists some positive integer $L_{0}$
such that when $L>L_{0}$ we have $\left\vert C-C^{\prime}\right\vert
<\epsilon$, where $C^{\prime}$ denotes the rate obtained by employing the
suboptimal power allocation scheme $\left\{  P_{n}^{\prime}\right\}  $.

Although we have demonstrated the asymptotic optimality of an arbitrary number
of power loading schemes, including statistical waterfilling, it is essential
from a practical viewpoint that we understand the convergence behavior of this
result. Convergence can be studied by observing the rate at which the upper
and lower bounds on capacity, given by (\ref{eq:ulb}), approach the limit in
$L$. This leads to our second main result.

\begin{proposition}
Statistical waterfilling converges to the optimal power loading strategy like
$\mathcal{O}\left(  1/\left(  L\log L\right)  \right)  $, whereas for more
general asymptotically optimal power loading strategies, convergence is, at
worst, $\mathcal{O}\left(  1/\log L\right)  $.
\end{proposition}

\begin{IEEEproof}
The proof of the first part follows by substituting $P_{n}^{\prime}=P_{n}^{s}$ in the expression for $r_L$ given by (\ref{eq:rL}) and expanding at $L=\infty$.  The second part follows from the proof of Proposition \ref{th:mrc_optimal_power}).
\end{IEEEproof}

From this proposition, we see that although there are arbitrarily many
asymptotically optimal power loading strategies, convergence to optimality is
guaranteed to be significantly faster for statistical waterfilling than for
general schemes. Consequently, statistical waterfilling is not only easily
implemented and intuitive, but is a practical solution for systems with
moderate levels of inherent diversity.

\section{Numerical Evaluation\label{sec:eval}}

One example of a parallel channel with subchannels that
experience unequal mean fading gains can be found in OFDM-based UWB systems.
In this section, we evaluate our findings by considering both measured and simulated UWB channels.

\subsection{Measured Channels\label{sec:measured}}

A channel measurement campaign was conducted using a state-of-the-art $2 \times 4$ time-domain multi-antenna UWB channel
sounder.  The sounder, manufactured by MEDAV, interrogates the propagation channel by
using trains of pseudo noise (PN) sequences of 4095 pulses or chips
\cite{Sachs2007a}. These are generated in baseband at a clock rate of 6.95~GHz
and later up-converted, using the same clock, to cover the bandwidth from
approximately 3.5 to 10.5~GHz. In turn, the receiver down-converts captured
signals, does a periodic sub-sampling of them, and uses a phase shifter to allow
the sampling of complex channel impulse responses (CIRs) in the time domain.

\begin{figure}[ptb]
\begin{center}
\includegraphics[
width=8.2cm
]{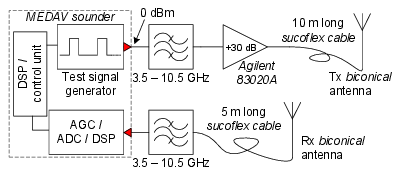}
\end{center}
\caption{Interconnection of sounding equipment.}%
\label{fig:sounder}%
\end{figure}

Fig.~\ref{fig:sounder} shows the configuration of the sounding equipment for
measurements. Test signals are generated and bandpass filtered to avoid
out-of-band emissions. The test signals are then amplified, transferred to a
biconical antenna for radiation and, after travelling through the propagation
environment, received by a similar antenna connected to a bandpass filter
\cite{IRK, cepedaIEE08}. After correcting the gain of the incoming signals
with an automatic gain control (AGC) unit, the receiver uses an
analogue-to-digital converter (ADC) and performs a matched filtering of the data,
with the known PN sequence, in a digital signal processing (DSP) unit. Before
sounding and after a warming-up period, the system response, phase imbalance
and crosstalk are characterized using cabled or open connections. These
measured parameters are then used for calibration to leave only the combined
response of the antennas and the environment on the data recordings.

UWB $(2\times4)$ MIMO channel measurements were conducted in an open-plan modern office 
in the central area of Bristol, UK. The sounding environment had typical
scattering objects such as personal computers, liquid crystal displays (LCDs),
non-metallic cubicle partitions, desks and metallic cabinets. The ceiling of
the office was made of perforated metallic tiles and the floor supported by a
metallic structure covered with non-metallic material. The dimensions of the sounded office are: 12.74~m
wide, 30.84~m long and 2.39~m high.

For this work, each recorded CIR results from averaging 256 captured CIRs in
hardware. Under these conditions, the observation time of the system for a
recorded CIR is 155~ms. The antennas were mounted on fibreglass masts at 1.3~m
from the floor. The transmit antenna mast was attached to the movable part of
an $x-y$ automated positioning system \cite{404XE}. The positioners and the
sounder were remotely controlled to prevent human
intervention in the area of measurements. In parallel, a spectrum analyzer,
connected to a biconical antenna and a low noise amplifier, periodically scanned
the spectrum of interest. This information was was used to check the \textquotedblleft health\textquotedblright%
\ of the test signal and the presence of interfering signals.

\begin{figure}[ptb]
\begin{center}
\includegraphics[
width=8.2cm
]{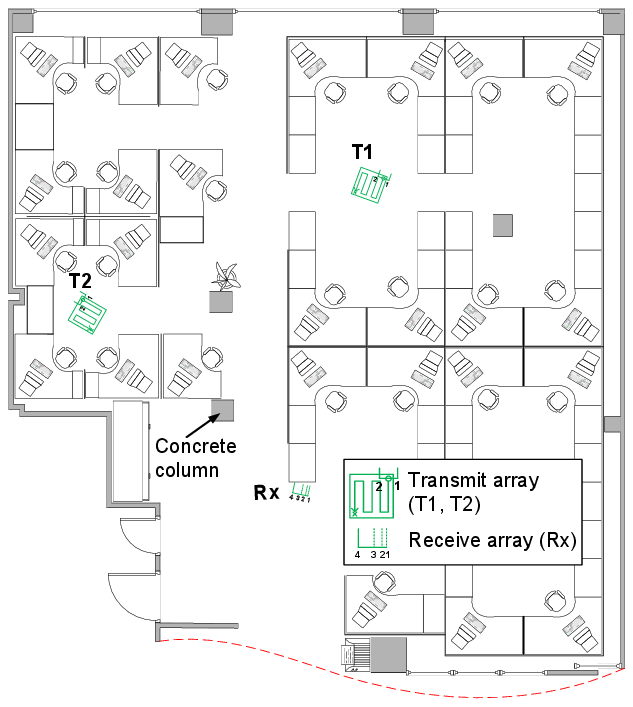}
\end{center}
\caption{Illustration of sounded environment and antenna locations.}%
\label{fig:plan32}%
\end{figure}

Fig.~\ref{fig:plan32} shows a diagram of the sounded environment. Two specific
locations were selected to capture line-of-sight (LOS) and non-line-of-sight
(NLOS) data. These locations, \textsf{T1} (LOS) and \textsf{T2} (NLOS), are
grids of $x-y$ points in which two transmit antennas, 25~cm apart, were displaced at distances
of 3~cm. The receive antenna mast was positioned at a fixed location (\textsf{Rx}), and a four-element linear
antenna array was formed by securing the first antenna and the following ones at 3, 6 and 12~cm from it. Each measurement grid had 441 $(21\times21)$ points, so a total of 3528 $(441\times8)$ CIRs
were recorded for each location \textsf{T}. In general, the distance between transmit and receive
antennas ranges from approximately 7.02~m to 6.51~m for \textsf{T1} and from
6.41~m to 5.77~m for \textsf{T2}. Note that NLOS conditions were achieved by
locating the transmit antenna in such a way that a concrete column was always
shadowing it from the receive one.

\begin{figure}
[ptb]
\begin{center}
\includegraphics[
width=8.2cm
]%
{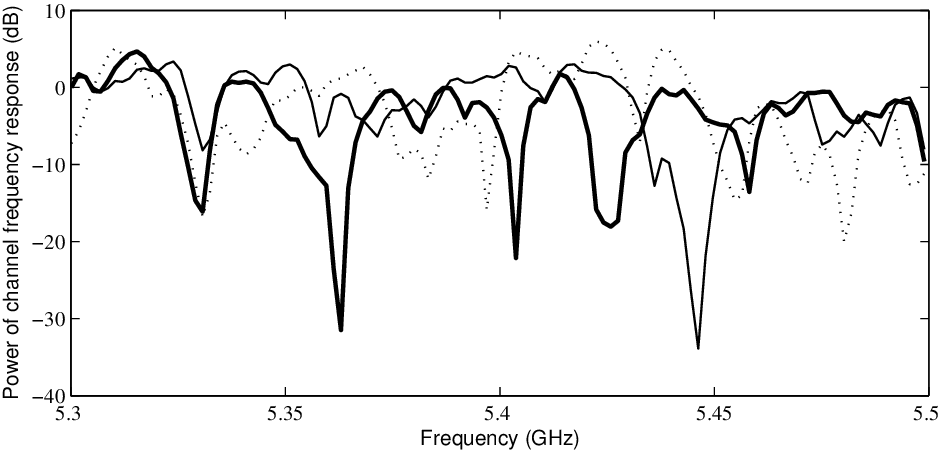}%
\caption{Plot of the power of three consecutive channel frequency response
measurements in the 5.3-to-5.5 GHz band.}%
\label{fig:meas_plot}%
\end{center}
\end{figure}

SIMO channels are formed by extracting the CIRs measured from the first transmit antenna to the receive antennas spaced apart by 12~cm distance. Each of these CIR is converted
into the frequency domain, by using a discrete Fourier transform (DFT), and we
retain the channel frequency response characteristics for the 5-to-6 GHz band.
With a frequency spacing of roughly 1.7 MHz, this amounts to 588 frequency
samples, i.e., channel frequency response coefficients. A power plot of three
consecutive NLOS measurements in the 5.3-to-5.5 GHz band is illustrated in
Fig. \ref{fig:meas_plot}. From this figure, we can see that the measurements
are fairly independent of one another.%

\begin{figure}
[ptb]
\begin{center}
\includegraphics[
width=8.2cm
]%
{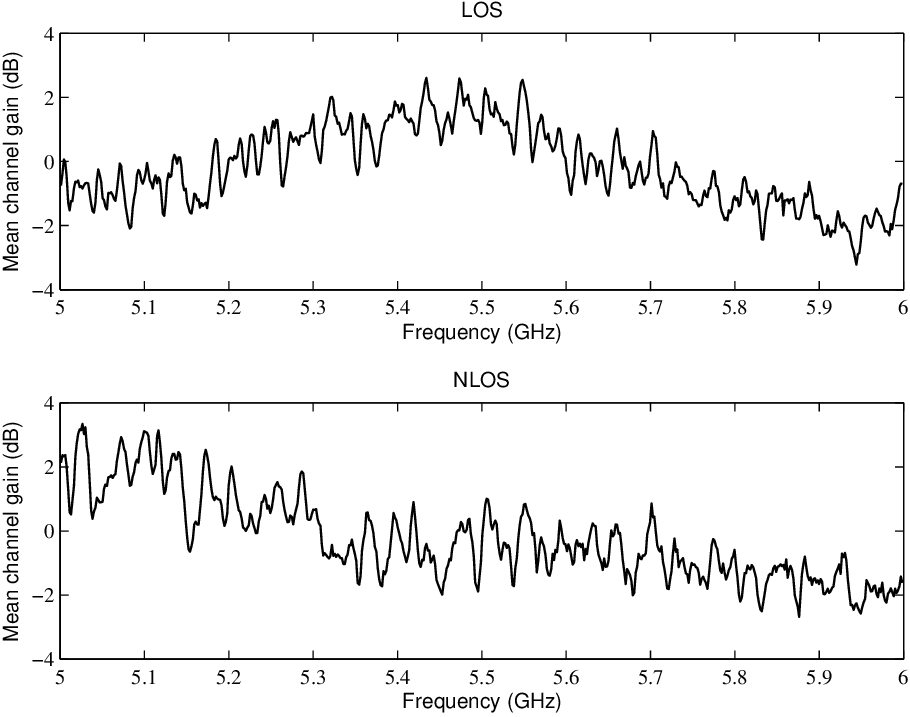}%
\caption{Mean channel gains for LOS and NLOS measured channels in the 5-to-6
GHz band.}%
\label{fig:mean_gains}%
\end{center}
\end{figure}

The channel measurements were normalized such that the average mean channel
gain (in frequency) is one. Fig. \ref{fig:mean_gains} depicts the mean channel
gain (averaged over the 441 available snapshots) for the LOS and NLOS
channels. It is certainly clear from this figure that the mean fading gains
vary considerably with frequency. Thus, one would expect an unequal power
loading strategy based on the statistics of these channels to perform better
than a balanced power allocation scheme.%

\begin{figure}
[ptb]
\begin{center}
\includegraphics[
width=8.2cm
]%
{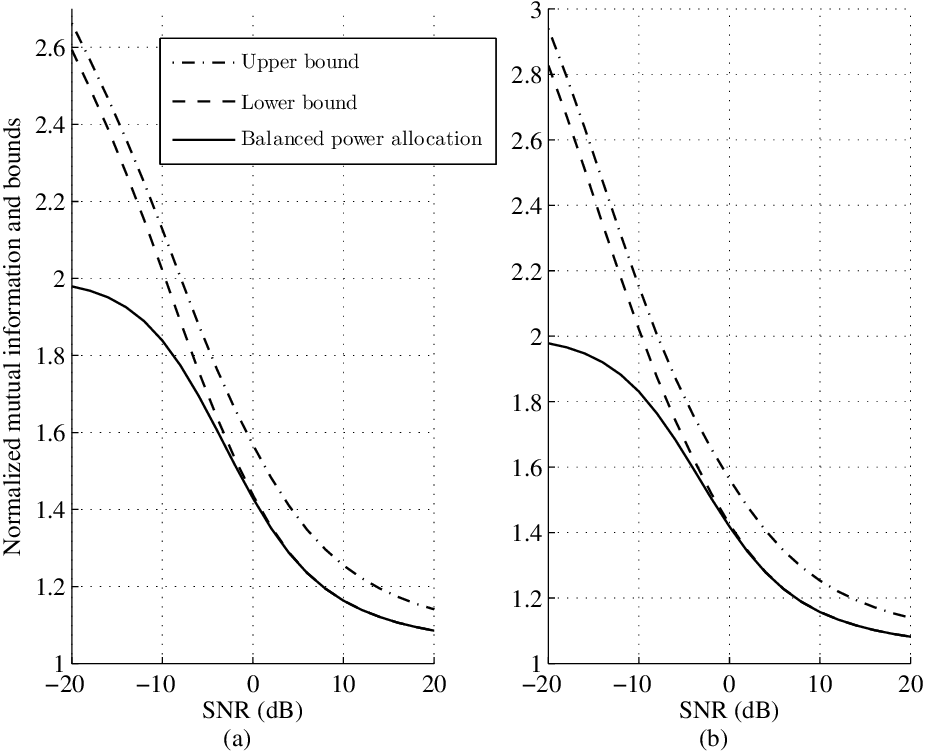}%
\caption{Normalized mutual information (w.r.t. a parallel AWGN channel) and bounds for measured LOS channels (a) and NLOS channels (b).}%
\label{fig:bounds}%
\end{center}
\end{figure}

In Fig. \ref{fig:bounds}, bounds on the capacity, defined by (\ref{eq:ub1}) and (\ref{eq:lb1}), are plotted for the measured
LOS and NLOS channels. Additionally, the mutual information for a channel where a balanced power allocation is used is illustrated.
The lower bound and the balanced power rate are
both averaged over the $441$ snapshots of measured data to emulate the
expectation in the rate expression. 
Since a large range of SNR
values are considered in these graphs, it is beneficial, for ease of comparison,
to normalize these results with respect to the capacity of a parallel
additive white Gaussian noise (AWGN) channel. It is evident from Fig. \ref{fig:bounds} that
statistical waterfilling is capable of providing significant gains at low SNR,
which exemplifies typical operating conditions in wideband systems with a
strict power budget, such as UWB.

\begin{figure}
[ptb]
\begin{center}
\includegraphics[
width=8.2cm
]%
{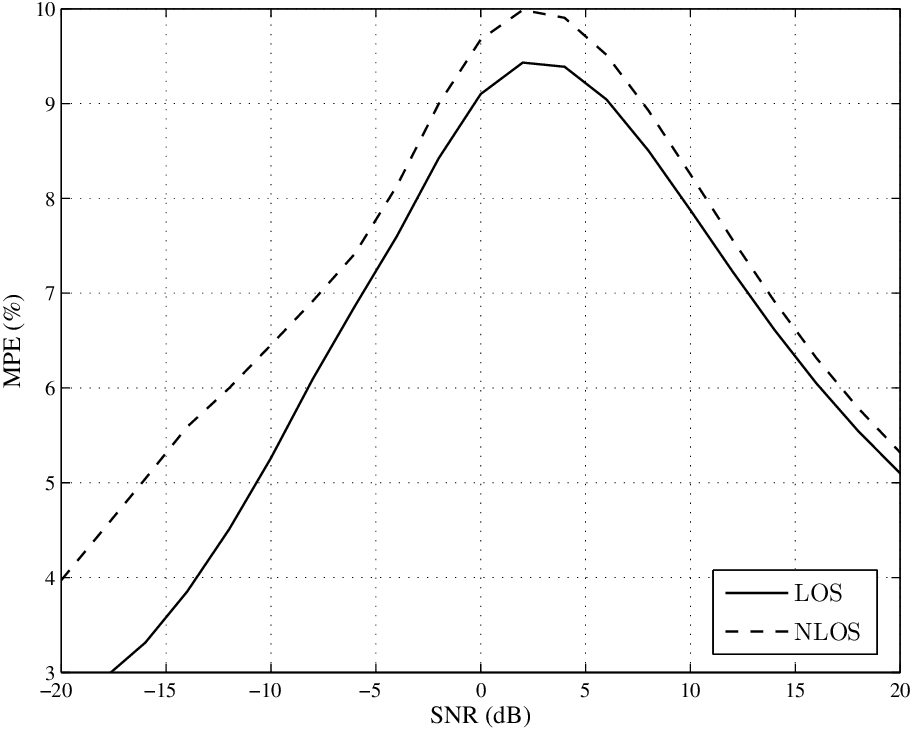}%
\caption{MPE for measured LOS and NLOS channels.}%
\label{fig:mpe}%
\end{center}
\end{figure}

It is also useful to examine the \emph{maximum percent error} (MPE) of the
capacity bounds, which is defined as $\mathsf{MPE}=100\% \times \left(C_{UB}-C_{LB}\right)/C_{LB}$
where $C_{UB}$ and $C_{LB}$ are the upper and lower bounds given by
(\ref{eq:ub1}) and (\ref{eq:lb1}), respectively. 
This metric quantifies the
deviation of the statistical waterfilling power allocation strategy from the
optimal power allocation since the power-optimal rate lies between
$C_{LB}$ and $C_{UB}$. 
The MPE is illustrated in Fig. \ref{fig:mpe}, which shows that even for the practical situation of $L=4$, the difference in the two bounds is relatively small, from which we deduce that statistical waterfilling is a good pragmatic power loading strategy.

\subsection{Simulated Channels\label{sec:sim}}
In the measured scenario, we are limited to studying the mutual information of a SIMO channel with $L \leq 4$ receive antennas.  Thus, we employ simulations to observe the performance of various power loading strategies as a function of $L$.  To this end, we consider a system operating in the 5-to-6 GHz band as above, but where the mean channel gain decays with frequency like $f^3$, which is typical in rich scattering environments.  

\begin{figure}
[ptb]
\begin{center}
\includegraphics[
width=8.2cm
]%
{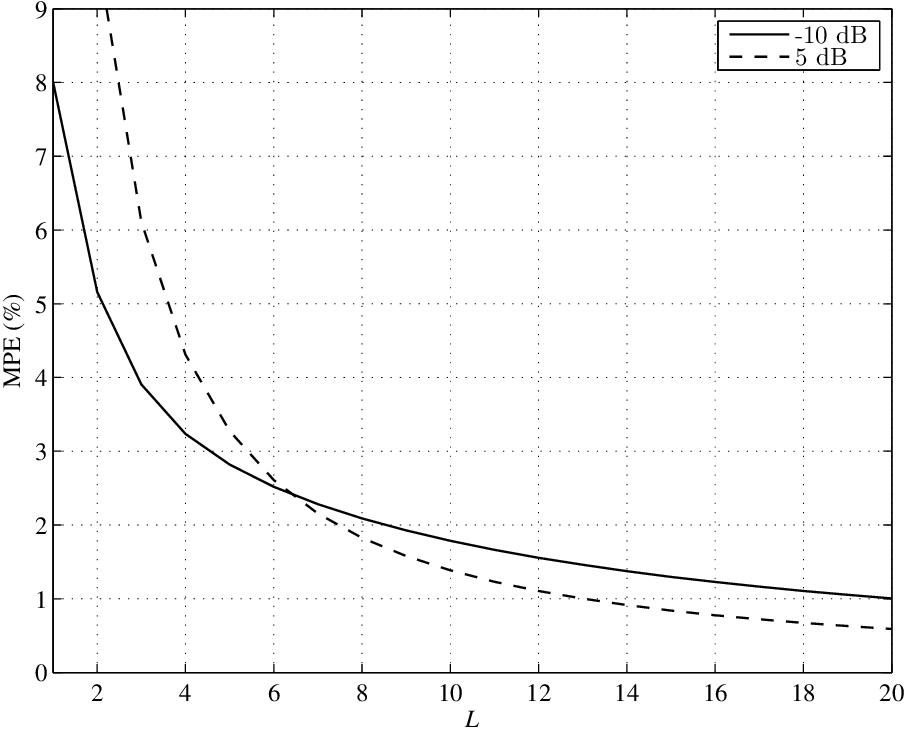}%
\caption{MPE for measured LOS and NLOS channels.}%
\label{fig:mpe_theory}%
\end{center}
\end{figure}

We are particularly interested in the MPE of the upper and lower capacity bounds, and plot this metric in Fig. \ref{fig:mpe_theory} as a function of $L$ for SNR values of $-10$ dB and $5$ dB.  This figure confirms that statistical waterfilling converges to the optimal power loading strategy quickly, achieving an MPE of less than $5\%$ for $L\geq4$, thus making this technique suitable for many practical systems.

\section{Conclusions\label{sec:conclusions}}

In this paper, we showed that there exist an arbitrary number of power loading
strategies that achieve capacity for parallel channels comprised of SIMO
subchannels in the limit of a large number of degrees of freedom $L$.
Statistical waterfilling was shown to be one such strategy, which in fact is
guaranteed to converge to the optimal solution much quicker than other
asymptotically optimal solutions. We demonstrated the practicality of our results through a study of measured and simulated UWB channels.



\bibliographystyle{IEEEtran}
\bibliography{acompat,IEEEabrv,master}

\end{document}